\documentclass[10pt,twocolumn,showpacs,longbibliography,amsmath,amssymb,osajnl,floatfix,superscriptaddress]{revtex4-1}

\usepackage{graphicx}
\usepackage{dcolumn}
\usepackage{bm}
\usepackage{color}
\usepackage{txfonts}
\usepackage{microtype}

\begin{document}

\author{Yan-Fei Li}	\affiliation{MOE Key Laboratory for Nonequilibrium Synthesis and Modulation of Condensed Matter, School of Science, Xi'an Jiaotong University, Xi'an 710049, China}\affiliation{Max-Planck-Institut f\"{u}r Kernphysik, Saupfercheckweg 1,
	69117 Heidelberg, Germany}
\author{Rashid Shaisultanov}
\affiliation{Max-Planck-Institut f\"{u}r Kernphysik, Saupfercheckweg 1,
	69117 Heidelberg, Germany}
\author{Yue-Yue Chen}
\affiliation{Max-Planck-Institut f\"{u}r Kernphysik, Saupfercheckweg 1,
	69117 Heidelberg, Germany}		
\author{Feng Wan}	\affiliation{MOE Key Laboratory for Nonequilibrium Synthesis and Modulation of Condensed Matter, School of Science, Xi'an Jiaotong University, Xi'an 710049, China}	\affiliation{Max-Planck-Institut f\"{u}r Kernphysik, Saupfercheckweg 1,
	69117 Heidelberg, Germany}
\author{Karen Z. Hatsagortsyan}\email{k.hatsagortsyan@mpi-hd.mpg.de}
\affiliation{Max-Planck-Institut f\"{u}r Kernphysik, Saupfercheckweg 1,	69117 Heidelberg, Germany}
\author{Christoph H. Keitel}
\affiliation{Max-Planck-Institut f\"{u}r Kernphysik, Saupfercheckweg 1,
	69117 Heidelberg, Germany}
\author{Jian-Xing Li}\email{jianxing@xjtu.edu.cn}
\affiliation{MOE Key Laboratory for Nonequilibrium Synthesis and Modulation of Condensed Matter, School of Science, Xi'an Jiaotong University, Xi'an 710049, China}\affiliation{Max-Planck-Institut f\"{u}r Kernphysik, Saupfercheckweg 1,
	69117 Heidelberg, Germany}

\title{Polarized ultrashort brilliant multi-GeV  $\gamma$-rays via  single-shot laser-electron interaction}

\date{\today}

\begin{abstract}

Generation of circularly-polarized (CP) and linearly-polarized  (LP) $\gamma$-rays via the single-shot interaction of an ultraintense laser pulse with a spin-polarized counterpropagating ultrarelativistic electron beam has been investigated in nonlinear Compton scattering in the quantum radiation-dominated regime. For the process simulation a Monte Carlo method is developed which employs the electron-spin-resolved probabilities for polarized photon emissions. We show efficient ways for the transfer of the electron polarization to the high-energy photon polarization.
In particular,  multi-GeV CP (LP) $\gamma$-rays with polarization of up to about 95\% can be generated  by a longitudinally (transversely) spin-polarized electron beam, with a photon flux meeting the requirements of recent proposals for the vacuum birefringence measurement in ultrastrong laser fields. Such high-energy, high-brilliance, high-polarization $\gamma$-rays are also beneficial for other applications  in high-energy physics,  and laboratory astrophysics.

\end{abstract}

\maketitle

Polarization is a crucial intrinsic property of a $\gamma$-photon. In astrophysics the $\gamma$-photon polarization provides detailed insight into the $\gamma$-ray emission mechanism 
and on properties of dark matter \cite{Laurent2011,Bohm_2017}. Highly-polarized high-energy $\gamma$-rays are a versatile tool in high-energy  \cite{Moortgat2008} and nuclear physics \cite{Uggerhoj2005}.  For instance, polarized $\gamma$-rays of tens of MeV can be used 
to excite polarization-dependent photofission of the nucleus in the giant dipole resonance \cite{Speth1981}, while polarized $\gamma$-rays of up to GeV play crucial roles for the meson-photoproduction \cite{Akbar2017}.

Recently, several proposals have been put forward to detect vacuum birefringence in ultrastrong laser fields, probing it with circularly-polarized (CP) or lineraly-polarized (LP) $\gamma$-photons of high-energies (larger than MeV and up to several GeV),
see \cite{King_2016,Ilderton_2016,Ataman_2017,Nakamiya_2017,Bragin2017},
taking advantage of the fact that the QED vacuum nonlinearity is significantly enhanced for high-energy photons. As proved in \cite{Bragin2017}, the use of CP rather than LP probe photons reduces the measurement time of vacuum birefringence and vacuum dichroism by two orders of magnitude.

The common ways of producing high-energy polarized $\gamma$-rays are linear Compton scattering \cite{Bocquet1997,Nakano1998,Omori_2006,Alexander_2008,Blanpied1999}
and bremsstrahlung  \cite{Maximon_1959,Kuraev_2010,Abbott_2016,Lohmann1994}. The advantage of the former is that it employs unpolarized electron beams, and the emitted $\gamma$-photon polarization is determined by the driving laser polarization, while in the latter the spin of the scattering electron determines the $\gamma$-photon polarization \cite{Berestetskii_1982}. However, in linear Compton scattering the electron-photon collision luminosity is rather low. The collision luminosity can be increased by using high-intensity lasers, but in this case the interaction regime moves into the nonlinear regime, when the radiation formation length is much smaller than the laser wavelength.   Then during the photon formation the laser field does not vary much and the emission process acquires similarity to bremsstrahlung. Namely, in the nonlinear regime the circular polarization of the emitted $\gamma$-photon requires  longitudinally spin-polarized (LSP) electrons. Nevertheless, the nonlinear regime of Compton scattering is  beneficial for the generation of polarized $\gamma$-photons, because the polarization is enhanced at high $\gamma$-photon energies \cite{Bocquet1997}, and  the  typical emitted photon energy is increased in the quantum nonlinear regime,  becoming comparable with the electron energy
\cite{Ritus_1985,Piazza2012}. 
 In the nonlinear regime the relative bandwidth of emission is increased \cite{Boca_2011}, which is not suitable for photonic applications involving narrow resonances, and stimulated investigations for the bandwidth reduction, see e.g. \cite{Ghebregziabher_2013,Terzi_2014}. However, vacuum birefringence is not a resonant effect and its measurement does not require a small $\gamma$-photon bandwidth, but mostly high flux of highly-polarized high-energy photons.

Regarding deficiencies connected with the bremsstrahlung mechanism, the incoherent bremsstrahlung cannot generate LP $\gamma$-photons, and the scattering angle and emission divergence are both relatively large \cite{Baier1998}. Furthermore, for  coherent bremsstrahlung \cite{Ter-Mikaelian_1972, Uggerhoj2005}, the current density of the impinging electrons and the radiation flux is limited by the damage threshold of the crystal material~\cite{Lohmann1994,Carrigan_1987,Biryukov_1997}.

With rapid developments of strong laser techniques,  stable (energy fluctuation $\sim$ 1\%) ultrashort ultraintense laser pulses can reach peak intensities of the scale of $10^{22}$ W/cm$^2$ with a duration of about tens of femtoseconds
 \cite{Danson2015,Gales_2018,ELI,Vulcan,Exawatt,CORELS}, opening new ways to generate high-energy high-flux $\gamma$-rays \cite{Sarri2014,Cole2018prx,Poder_2018,Li2015,Magnusson_2018, Xie_2017} in the nonlinear regime of  Compton scattering \cite{Goldman_1964, Nikishov_1964, Ritus_1985,Bula_1996}. Moderately polarized  $\gamma$-photons have been predicted in strong fields in  electron-spin-averaged treatment \cite{King_2013,King_2016, Naveen_2019}. However, the polarization properties of radiation are essentially spin-dependent in the nonlinear regime \cite{Baier1998, Ivanov_2004}, differing from those in the linear regime
 \cite{CAIN, Sun_2011, Petrillo_2015, An_2018}), which calls for comprehensive spin-resolved studies, especially in the most attractive high-energy regime.

  \begin{figure}[t]
 	\includegraphics[width=0.8\linewidth]{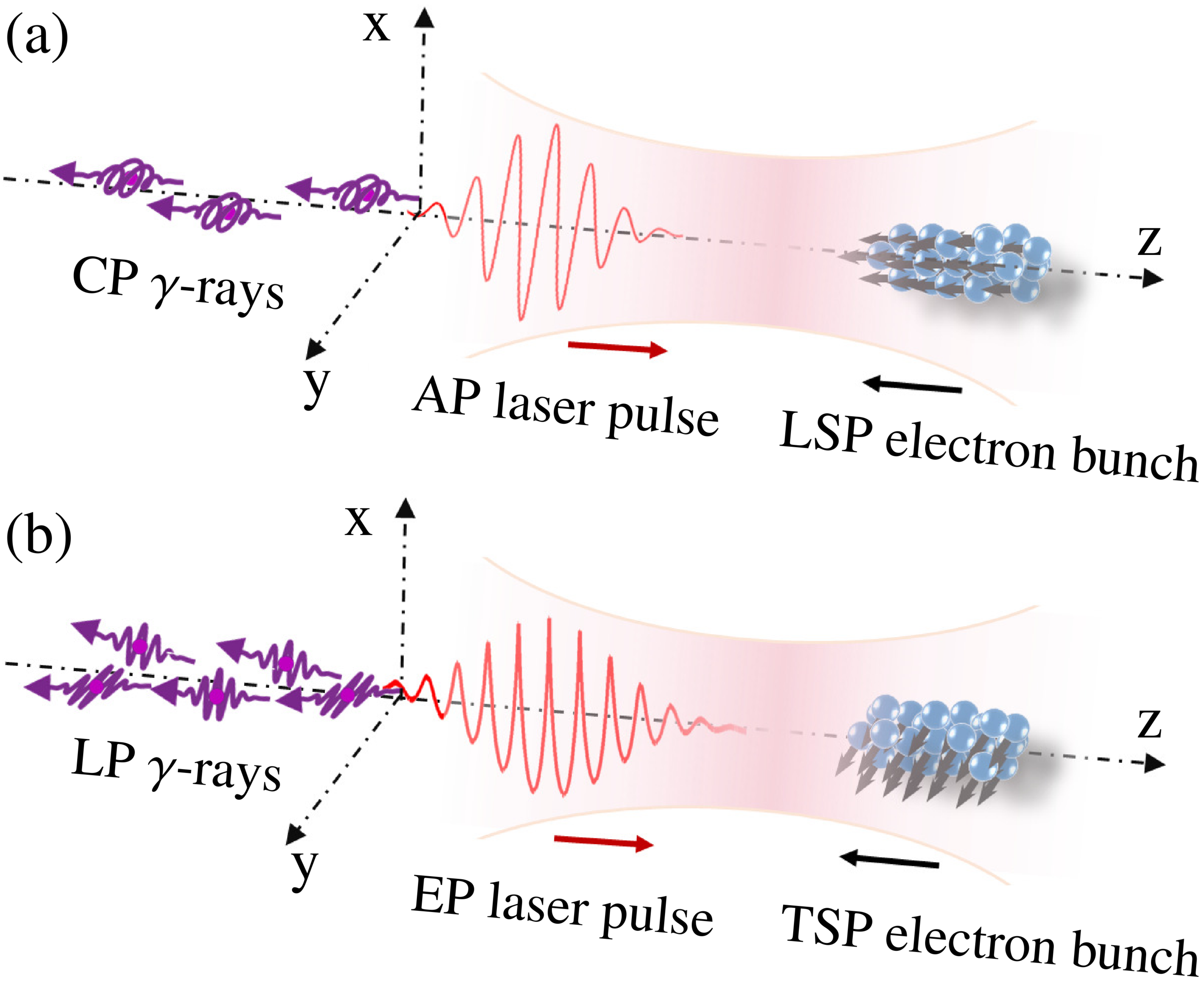}
 \caption{Scenarios of generating CP and LP $\gamma$-rays via nonlinear Compton scattering. (a): An arbitrarily-polarized (AP) laser pulse propagating along $+z$ direction and head-on colliding with a longitudinally spin-polarized (LSP) electron bunch produces CP $\gamma$-rays. (b): An elliptically-polarized (EP) laser pulse propagating along $+z$ direction and colliding with a transversely spin-polarized (TSP) electron bunch produces LP $\gamma$-rays.
 	The major axis of the polarization ellipse is along the $x$-axis.} \label{fig1}
 \end{figure}

In this Letter, the feasibility of generation of polarized  ultrashort multi-GeV brilliant $\gamma$-rays via nonlinear Compton scattering with spin-polarized electrons is investigated theoretically (see  Fig.~\ref{fig1}). High-flux  $\gamma$-photons with polarization beyond  95\%  are shown to be feasible in a single-shot interaction, along with new applications in high-energy, astro- and strong laser physics. The investigation is based on the developed Monte Carlo method for simulation of polarization-resolved radiative processes in the interaction of an ultrastrong laser beam  with a relativistic spin-polarized electron beam. While the scheme for CP $\gamma$-photons includes an arbitrarily-polarized (AP)  laser pulse colliding with a LSP electron bunch (see Fig.~\ref{fig1}(a)), the scheme for LP $\gamma$-photons employs an elliptically-polarized (EP) laser pulse with a small ellipticity colliding with a transversely spin-polarized (TSP) electron bunch (see Fig.~\ref{fig1}(b)). The spin-dependent radiation reaction in a laser field with a small ellipticity yields  the separation of $\gamma$-photons with respect to the polarization  and  the enhancement of the polarization rate.

Let us first introduce our new Monte Carlo method for simulation of polarized $\gamma$-photon emissions during the interaction of polarized relativistic electrons with ultrastrong laser fields. Photon emissions are treated quantum mechanically, while the electron dynamics semiclassically. At each simulation step the photon emission is determined by the total emission probability, and the photon energy by the spectral probability, using the common algorithms ~\cite{Ridgers_2014,Elkina2011,Green2015}. The spin of the electron after the emission is determined by the spin-resolved emission probabilities according to the algorithm of Ref.~\cite{li2019prl}; See more details in \cite{supplemental}. For determination of the photon polarization, we employ the polarized photon emission probabilities by polarized electrons in the local constant field approximation \cite{Ritus_1985,Baier1998, piazza2018pra,Ilderton2019pra,Podszus2019prd,Ilderton2019prd, piazza2019}, which are derived in the QED operator method of Baier-Katkov \cite{Baier_1973}.  This approximation is valid in ultrastrong laser field, when the invariant  field parameter is large $a_0\equiv |e|E_0/(m\omega_0)\gg 1$   \cite{Ritus_1985, Baier1998}, with  laser field amplitude $E_0$, frequency $\omega_0$, and the electron charge $e$ and mass $m$.
Relativistic units with $c=\hbar=1$ are used throughout.
 The radiation probabilities are characterized by the quantum strong field parameter  $\chi\equiv |e|\sqrt{-(F_{\mu\nu}p^{\nu})^2}/m^3$ \cite{Ritus_1985}, where  $F_{\mu\nu}$ is the field tensor, and $p^{\nu}$ the four-vector of the electron momentum. The angle-integrated radiation probability of a polarized photon with a polarized electron reads:
\begin{equation}\label{W}
\frac{{\rm d^2}W_{fi}}{{\rm d}u{\rm d}\eta}=\frac{W_R}{2}\left(F_0+\xi_1 F_1+\xi_2 F_2 + \xi_3 F_3\right),
\end{equation}
where $W_R={\alpha m}/\left[{8\sqrt{3}\pi\lambdabar_c\left( k\cdot p_i\right)}{\left(1+u\right)^3}\right]$, $\alpha$ is the fine structure constant, $u=\varepsilon_\gamma/\left(\varepsilon_i-\varepsilon_\gamma\right)$,  $\lambdabar_c$  the Compton wavelength, $\varepsilon_\gamma$ the emitted photon energy, $\varepsilon_i$ the electron energy before radiation,  $\eta=k\cdot r$ the laser phase, while $p_i$, $k$, and $r$  are  four-vectors of the electron momentum  before radiation, laser wave-vector, and coordinate, respectively. The photon polarization is represented by the Stokes parameters  ($\xi_1$, $\xi_2$, $\xi_3$), defined with respect to the axes $\hat{{\bf e}}_1=\hat{\bf a}-\hat{\bf v}(\hat{\bf v}\hat{\bf a})$ and $\hat{{\bf e}}_2=\hat{\bf v}\times\hat{\bf a}$ \cite{McMaster_1961}, with the photon emission direction $\hat{\bf n}$ along the electron velocity ${\bf v}$ for the ultrarelativistic electron, $\hat{\bf v}={\bf v}/|{\bf v}|$,
and the unit vector $\hat{{\bf a}}={\bf a}/|{\bf a}|$ along the electron acceleration  ${\bf a}$. The variables introduced in Eq.~(\ref{W}) read:
\begin{eqnarray}
\label{F0}
F_0&=&-(2+u)^2 \left[{\rm IntK}_{\frac{1}{3}}(u')
 -2{\rm K}_{\frac{2}{3}}(u') \right](1+{\bf S}_{if})+u^2(1-{\bf S}_{if})\nonumber\\
 &&\left[{\rm IntK}_{\frac{1}{3}}(u')
+2{\rm K}_{\frac{2}{3}}(u') \right]+2u^2{\bf S}_{if}{\rm IntK}_{\frac{1}{3}}(u')-(4u+2u^2)\nonumber\\
&&({\bf S}_f+{\bf S}_i)\left[\hat{\bf v}\times\hat{{\bf a}}\right]{\rm K}_{\frac{1}{3}}(u')-2u^2({\bf S}_f-{\bf S}_i)\left[\hat{\bf v}\times\hat{{\bf a}}\right]{\rm K}_{\frac{1}{3}}(u')\nonumber\\
&&-4u^2\left[{\rm IntK}_{\frac{1}{3}}(u')
-{\rm K}_{\frac{2}{3}}(u') \right]({\bf S}_i\cdot\hat{\bf v})({\bf S}_f\cdot\hat{\bf v}),\\
\label{F1}
F_1&=&-2u^2{\rm IntK}_{\frac{1}{3}}(u')
\left\{({\bf S}_{i}\hat{\bf a}){\bf S}_{f}\left[\hat{\bf v}\times\hat{\bf a}\right]+({\bf S}_{f}\hat{\bf a}){\bf S}_{i}\left[\hat{\bf v}\times\hat{\bf a}\right]\right\}+\nonumber\\
&&4u\left[({\bf S}_i\cdot\hat{\bf a})(1+u)+({\bf S}_f\cdot\hat{\bf a})\right]{\rm K}_{\frac{1}{3}}(u')+\nonumber\\
&&2u(2+u)\hat{\bf v}[{\bf S}_f\times{\bf S}_i]{\rm K}_{\frac{2}{3}}(u'),\\
\label{F2}
F_2&=&-\left\{2u^2 \left\{({\bf S}_{i}\hat{\bf v}){\bf S}_{f}\left[\hat{\bf v}\times\hat{\bf a}\right]+({\bf S}_{f}\hat{\bf v}){\bf S}_{i}\left[\hat{\bf v}\times\hat{\bf a}\right]\right\}+2u(2+u)\right.\nonumber\\
&&\left.\hat{\bf a}[{\bf S}_f\times{\bf S}_i]\right\}{\rm K}_{\frac{1}{3}}(u')-4u\left[({\bf S}_i\cdot\hat{\bf v})+({\bf S}_f\cdot\hat{\bf v})(1+u)\right]\nonumber\\&&{\rm IntK}_{\frac{1}{3}}(u')+4u(2+u)\left[({\bf S}_i\cdot\hat{\bf v})+({\bf S}_f\cdot\hat{\bf v})\right]{\rm K}_{\frac{2}{3}}(u'),\\
\label{F3}
F_3&=&4\left[1+u+(1+u+\frac{u^2}{2}){\bf S}_{if}-\frac{u^2}{2}({\bf S}_i\cdot\hat{\bf v})({\bf S}_f\cdot\hat{\bf v})\right]{\rm K}_{\frac{2}{3}}(u')\nonumber\\
&&+2u^2\left\{{\bf S}_i\left[\hat{\bf v}\times\hat{\bf a}\right]{\bf S}_f\left[\hat{\bf v}\times\hat{\bf a}\right]-({\bf S}_i\cdot\hat{\bf a})({\bf S}_f\cdot\hat{\bf a})\right\}{\rm IntK}_{\frac{1}{3}}(u')\nonumber\\
&&-4u\left[(1+u){\bf S}_i\left[\hat{\bf v}\times\hat{\bf a}\right]+{\bf S}_f\left[\hat{\bf v}\times\hat{\bf a}\right]\right]{\rm K}_{\frac{1}{3}}(u'),
\end{eqnarray}
where $u'=2u/3\chi$, ${\rm IntK}_{\frac{1}{3}}(u')\equiv \int_{u'}^{\infty} {\rm d}z {\rm K}_{\frac{1}{3}}(z)$,  ${\rm K}_n$ is the $n$-order modified Bessel function of the second kind, ${\bf S}_{i}$ and ${\bf S}_f$ are the electron spin-polarization vector before and after radiation, respectively, $|{\bf S}_{i,f}|=1$, and ${\bf S}_{if}\equiv {\bf S}_i\cdot{\bf S}_f$.
Our Monte Carlo algorithm to determine the photon polarization yields the following, taking into account that the averaged polarization of the emitted photon is in a mixed state:
\begin{equation}
\label{average}
\xi_{1}^{mix}=F_1/F_0, \quad \xi_{2}^{mix}=F_2/F_0, \quad \xi_{3}^{mix}=F_3/F_0.
\end{equation}
Here we choose as the basis for the emitted photon \cite{Lipps_1954, McMaster_1961}
the two orthogonal pure states  with the Stokes parameters 
$\hat{\boldsymbol{\xi}}^{\pm}\equiv\pm$($\xi_1^{mix}$, $\xi_2^{mix}$, $\xi_3^{mix}$)/$\xi_0^{mix}$ with $\xi_0^{mix}\equiv \sqrt{(\xi_1^{mix})^2+(\xi_2^{mix})^2+(\xi_3^{mix})^2}$. 

\begin{figure}[t]
\setlength{\abovecaptionskip}{-0cm}  	
	\includegraphics[width=1\linewidth]{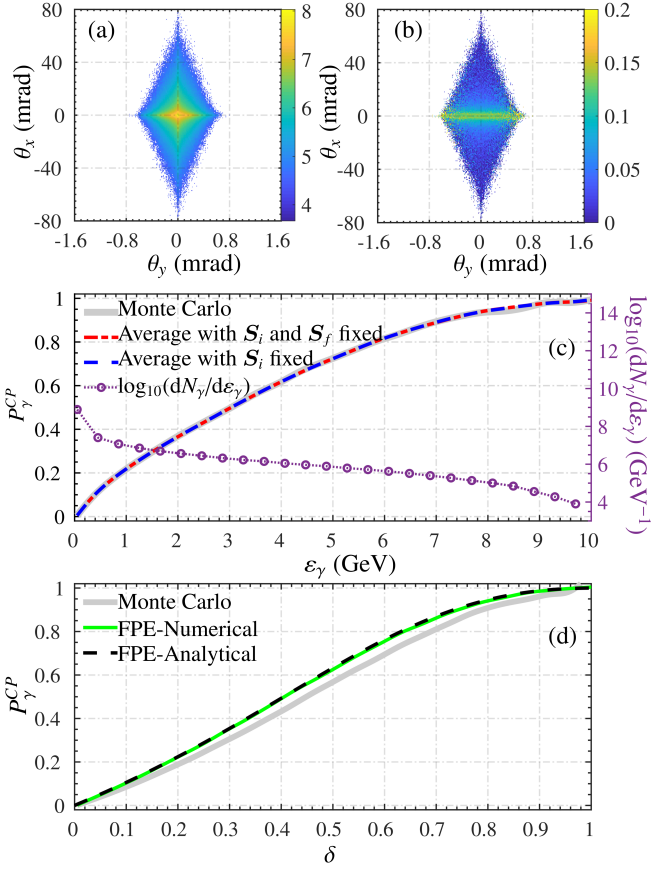}
		\caption{(a) Angle-resolved photon number density log$_{10}$(d$^2N_p$/d$\theta_x$d$\theta_y$) (mrad$^{-2}$) vs deflection angles, $\theta_x=p_x/p_z$ and $\theta_y=p_y/p_z$; (b) Distribution of average  circular polarization for all emitted photons $\overline{P^{CP}}=\overline{\xi_2}$;
		(c) Degree of circular polarization of $\gamma$-photons $P_\gamma^{CP}=\xi_2$ and energy density of $\gamma$-photons log$_{10}$(d$N_\gamma$/d$\varepsilon_\gamma$) vs $\gamma$-photon energy $\varepsilon_\gamma$:  via the Monte Carlo method (gray-thick-solid),  via the spin-resolved average method of Eq.~(\ref{average}) (red-dash-dotted), and  via the average method summed up by $\bm S_f$ at each simulation step (blue-dashed), respectively. (d)  $P_\gamma^{CP}$  vs the energy ratio parameter $\delta=\varepsilon_\gamma/\varepsilon_e$: via the Monte Carlo method (gray-thick-solid),
and		only considering the first photon emission (FPE) calculated  numerically (green-thin-solid, using the instantaneous  $\chi$ parameter) and analytically (black-dashed, employing the constant average parameter of $\chi=1.41$) via the average method (summing over $\bm S_f$), respectively.  In (b) the observation frame can be chosen following \cite{supplemental}.  In (c) and (d)  the basis vectors of the observation frame  $\hat{{\textbf o}}_1$, $\hat{{\textbf o}}_2$ and $\hat{\textbf o}_3$ are along the $x$ axis, $-y$ axis and $-z$ axis, respectively. The laser and other electron beam parameters are given in the text.}
		\label{fig2}
\end{figure}

Using the probabilities for the photon emission in these states $W_{fi}^\pm$ given by Eq.~(\ref{W}), we define the stochastic procedure with a random number $N_r\in [0, 1]$:  if $W_{fi}^+/\overline{W}_{fi}\geq N_r$, the $\hat{\boldsymbol{\xi}}^{+}$ photon state is chosen, otherwise the photon state is set to $\hat{\boldsymbol{\xi}}^{-}$, see \cite{supplemental}. 
The Stokes parameters of each emitted photon are rotated from the instantaneous frame $(\xi_1,\xi_2,\xi_3)$ defined with respect to the basis vectors $\hat{\textbf{e}}_1$, $\hat{\textbf{e}}_2$ and $\hat{\textbf{n}}$ to the observation frame $(\xi^{(o)}_1,\xi^{(o)}_2,\xi^{(o)}_3)$ defined with respect to the basis vectors $\hat{\textbf{o}}_1$, $\hat{\textbf{o}}_2$ and $\hat{\textbf{o}}_3$, see \cite{supplemental, McMaster_1961}.
Here, $\overline{W}_{fi}\equiv W_R F_0$ is the electron-spin-resolved radiation probability averaged by the photon polarization, cf.~\cite{li2019prl}. Averaging over the electron spin in Eq.~(\ref{W}) yields $\xi_{2}=0$, indicating that only  LP $\gamma$-photons can be generated with unpolarized electrons, in accordance with \cite{King_2013}.

The simulation results for the generation of CP $\gamma$-rays are shown in Fig.~\ref{fig2}.
A realistic tightly-focused Gaussian LP laser pulse is used \cite{Salamin2002, Salamin2002PhysRevSTAB, supplemental}, with  peak intensity  $I_0\approx3.45\times10^{21}$ W/cm$^2$ ($a_0=50$),  wavelength $\lambda_0=1$ $\mu$m,  pulse duration $\tau = 10T_0$ with  period $T_0$, and  focal radius $w_0=5$ $\mu$m.  The electron beam counterpropagating with the laser pulse (polar angle $\theta_e=180^{\circ}$ and azimuthal angle $\phi_e=0^\circ$) is fully LSP  with $(\overline{S}_x, \overline{S}_y, \overline{S}_z) = (0, 0, -1)$. It has a cylindrical form
with  radius  $w_e= 2\lambda_0$,  length $L_e = 3\lambda_0$ and electron number $N_e=5\times10^6$ (density $n_e\approx 1.33\times10^{17}$ cm$^{-3}$). The electron density has
a transversely Gaussian and longitudinally uniform distribution.
The electron initial kinetic energy is $\varepsilon_0=10$ GeV, the angular divergence $ \Delta\theta = 0.2$ mrad, and the energy spread $\Delta \varepsilon_0/\varepsilon_0 =0.06$.  The emittance of the electron beam is estimated  $\epsilon_e\approx 4\times10^{-4}$  mm$\cdot$mrad. In our simulations, the electron-positron pair creations and their further radiations are taken into account.
 Such electron bunches are achievable via laser wakefield acceleration \cite{Leemans2014,Leemans_2019} with further radiative polarization \cite{li2019prl},
 or alternatively, via directly wakefield acceleration of LSP electrons \cite{Wen_2018}.

The angle-resolved number density and average circular polarization of all emitted photons are given in Figs.~\ref{fig2}(a)~ and~(b).  The $\gamma$-ray pulse duration is determined by the electron bunch length: $\tau_\gamma\approx \tau_e\approx 10$~fs~\cite{Li2015}. The total number of emitted $\gamma$-photons $N_\gamma$ at these parameters is about $8.04\times 10^7$ for $\varepsilon_\gamma \geq 1$ MeV and about $1.23\times 10^7$ for $\varepsilon_\gamma \geq 1$ GeV, i.e., $N_\gamma$ is approximately one order of magnitude larger than $N_e$, which is in accordance with the analytical estimation $ N_\gamma  \sim \alpha a_0 N_e\tau/ T_0$ \cite{Ritus_1985}. For comparison the polarized $\gamma$-photon number in the linear Compton scattering is $N_\gamma/N_e\approx10^{-3}$ at $\varepsilon_\gamma\approx 56$ MeV \cite{Omori_2006}. The total flux of the $\gamma$-rays with $\varepsilon_\gamma \geq 1$ MeV is large due to the shortness of the pulse: ${\cal F}_\gamma\approx 8.04\times 10^{21}$~s$^{-1}$. The circular polarization of $\gamma$-photon $P_\gamma^{CP}$ is proportional to the $\gamma$-photon energy $\varepsilon_\gamma$ (derived from Eqs.~(\ref{F0}) and (\ref{F2})), and significantly higher for higher photon energies, see Fig.~\ref{fig2}(c).  Intuitively speaking, the polarization of emitted $\gamma$-photons (photon helicity in the case of CP) is transfered from the angular momentum  (helicity) of electrons. The larger the average energy of emitted $\gamma$-photons, the smaller the number of emitted photons per electron will be. As the electron carries a certain  helicity, the  transferred average helicity per photon will be larger in the case of smaller photon number, i.e., in the case of high photon energies.
For instance, multi-GeV $\gamma$-rays of about 10-GeV, 8-GeV and 6-GeV can be emitted with circular polarization of about 99\%, 94\% and 81\%, respectively, and with quite significant brilliances of about  $6.25\times10^{18}$, $3.94\times10^{20}$ and $1.36\times10^{21}$ photons/(s mm$^2$ mrad$^2$ 0.1\% BW), respectively, which are comparable with the brilliance of unpolarized multi-MeV $\gamma$-rays obtained in a recent experiment \cite{Sarri2014}.  The detection of vacuum  birefringence demands $N_\gamma\sim 10^7-10^9$ CP GeV $\gamma$-photons \cite{Bragin2017}. Taking into account that LSP  electron bunches with $N_e\approx 5\times 10^8$ are feasible \cite{Wen_2018}, we estimate the number of CP $\gamma$-photons at 4 GeV from Fig.~\ref{fig2}(c) for such bunches (using $\Delta \varepsilon_\gamma/\varepsilon_\gamma\sim 0.01$) to reach the required target number for vacuum  birefringence $N_\gamma\approx 10^6$ in a single-shot interaction.

Figure~\ref{fig2} indicates that
the photon polarization is mostly determined by the incoming electron spin vector  $\bm S_i$, where the Monte Carlo and the average methods provide identical results; See Fig.~\ref{fig2}(c).
During multiple photon emissions the average spin of the electron beam from the initial longitudinal direction is gradually oriented along the laser magnetic field in the transverse direction. This effect reduces  the circular polarization in the case of multiple photon emissions; See
Fig.~\ref{fig2}(d).

\begin{figure}[t]
	\setlength{\abovecaptionskip}{-0cm}  	
	\includegraphics[width=1\linewidth]{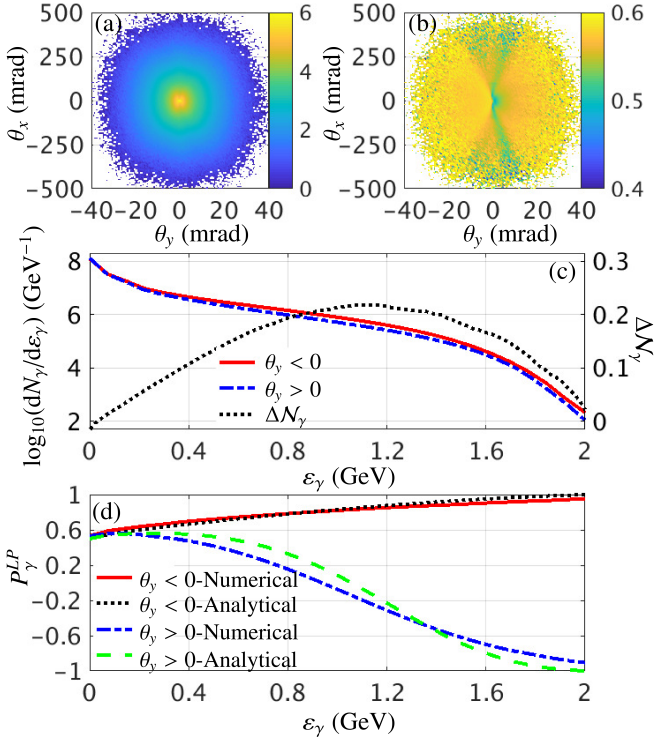}
	\caption{ LP $\gamma$-ray emission with a left-handed EP laser with $a_0=100$, $\tau=5T_0$, ellipticity $\epsilon=|E_y|/|E_x|=0.05$, $\varepsilon_0=2$ GeV,  initial average spin of electrons $(\overline{S}_x, \overline{S}_y, \overline{S}_z) = (0, 1, 0)$ with  100\% polarization, and other laser and electron beam parameters are the same as those in Fig.~\ref{fig2}: (a)  log$_{10}$(d$^2N_p$/d$\theta_x$d$\theta_y$) (mrad$^{-2}$) vs $\theta_x$ and $\theta_y$; (b) Average linear polarization for all emitted photons $\overline{P^{LP}}=\sqrt{\xi_1^2+\xi_3^2}$ vs $\theta_x$ and $\theta_y$; (c) log$_{10}$(d$N_\gamma$/d$\varepsilon_\gamma$) into -10 mrad $<\theta_y<0$ (red-solid) and into $0<\theta_y<$ 10 mrad (blue-dash-dotted), respectively, and the relative difference $\Delta \mathcal{N_\gamma}\equiv(dN_\gamma^{\theta_y<0}$/d$\varepsilon_\gamma$-d$N_\gamma^{\theta_y>0}$/d$\varepsilon_\gamma$)/(d$N_\gamma^{\theta_y<0}$/d$\varepsilon_\gamma$+d$N_\gamma^{\theta_y>0}$/d$\varepsilon_\gamma$) (black-dotted)  vs $\varepsilon_\gamma$. (d) Linear polarization of $\gamma$-photons $P_\gamma^{LP}$ (employing $\xi_3$)
	vs $\varepsilon_\gamma$. In (c) and (d), -10 mrad $<\theta_x<$ 10 mrad.
	The red-solid (blue-dash-dotted) and black-dotted (green-dashed) curves indicate the results of -10 mrad $<\theta_y<0$ ($0<\theta_y<10$ mrad), calculated numerically and analytically (with average $\chi\approx0.7$), respectively. In (d)  the basis vectors of the observation frame  $\hat{{\textbf o}}_1$, $\hat{{\textbf o}}_2$ and $\hat{\textbf o}_3$ are along the $x$ axis, $-y$ axis and $-z$ axis, respectively.} 
	\label{fig3}
\end{figure}

Generation of LP $\gamma$-rays is analyzed in Fig.~\ref{fig3}. As an unpolarized (or TSP) electron beam head-on collides with a LP laser pulse, an average polarization of about 55\% can be obtained \cite{supplemental}, pointed out already in \cite{King_2013, King_2016}. However,  by harnessing the scheme with an  EP laser pulse \cite{li2019prl},
much higher polarization can be achieved (see Fig.~\ref{fig1}(b)). We focus on the characteristics of the LP $\gamma$-photons in the high-density region of -10~mrad~$<\theta_{x,y}<$~10~mrad (in this region the total radiation flux ${\cal F}_\gamma\approx 2.82\times10^{21}$ s$^{-1}$ and the average linear polarization $\overline{P^{LP}}\approx$ 55.2\%).

Due to the electron-spin-dependence of radiation, namely, that the radiation probability is larger when the electron spin is anti-parallel to the rest frame magnetic field
\cite{li2019prl},
the TSP electrons more probably emit photons in the half cycles with $B_y<0$. In the given EP laser field, the $y$-component of the electron momentum  in those half-cycles with $B_y<0$ is positive, $p_y>0$, and the corresponding $\theta_y=p_y/p_z<0$, since $p_z<0$.
Then, the $\gamma$-photons are more emitted with $\theta_y<0$,
see Fig.~\ref{fig3}(c). The relative asymmetry of the photon emission $\Delta \mathcal{N_\gamma}$
corresponds to the relative difference of the radiation probabilities for ${\bm S_i}$ being parallel and anti-parallel to the magnetic field \cite{li2019prl} and is most significant around $\varepsilon_\gamma\approx1.2$ GeV ($\varepsilon_\gamma/\varepsilon_0\approx0.6$).
Due to radiative electron-spin effects, electrons and $\gamma$-photons are split into two parts along the  minor axis ($y$ axis) of the polarization ellipse  \cite{li2019prl}.
In contrast to the case with unpolarized electrons
\cite{King_2013, King_2016}, here at emission angles $\theta_y<0$ ($\theta_y>0$)  the $\gamma$-photon polarization is proportional (inversely proportional) to its energy, see Fig.~\ref{fig3}(d). Thus, specially highly LP $\gamma$-rays can be obtained in the high-energy region.

For $\theta_y<0$ in Fig.~\ref{fig3}(d), although the average linear polarization of all emitted photons is $\overline{P^{LP}}\approx$ 58.3\%, the high-energy photons achieve even higher linear polarization: at photon energies of about 2 GeV, 1 GeV, 0.4 GeV and 0.2 GeV,  the polarizations are of about 95\%, 82\%, 70\% and 64\%, respectively, and the brilliances of about $8.0\times10^{16}$, $1.2\times10^{20}$, $2.9\times10^{20}$ and $3.5\times10^{20}$ photons/(s mm$^2$ mrad$^2$ 0.1\% BW), respectively.  For $\theta_y>0$, the sign of polarization is energy dependent \cite{supplemental}.
Moreover, the depolarization effect due to multiple photon emissions
in this case is not significant, because subsequent photon emissions generate LP $\gamma$-photons as well, see the black-dotted and green-dashed curves in Fig.~\ref{fig3}(d).
Furthermore, using  $N_e\approx3\times10^8$  \cite{Wen_2018}, LP $\gamma$-photons of $N_\gamma\sim10^5-10^6$ within 1.015 GeV $\leq\varepsilon_\gamma\leq1.021$ GeV ($\Delta \varepsilon_\gamma/\varepsilon_\gamma\approx 0.0059$) can be obtained, which satisfy the requirement ($N_\gamma\gtrsim 6.4\times 10^4$) of the vacuum birefringence measurement with LP  photons \cite{Nakamiya_2017}.

Finally, we have investigated the impact of the laser and electron beam parameters on the polarization of $\gamma$-rays \cite{supplemental}. Generally, larger energy spread $\Delta \varepsilon_0/\varepsilon_0=0.1$, larger angular divergence of 1 mrad, and different collision angles $\theta_e=179^\circ$ and $\phi_e=90^\circ$ do not disturb significantly  the quality of $\gamma$-ray polarization. In generation of CP and LP $\gamma$-rays, the polarization is robust with respect to the variation of parameters, e.g., $\tau$, $\varepsilon_0$, $a_0$ and $\epsilon$. As the initial polarization of the electron beam decreases, the polarization of emitted $\gamma$-photons declines as well.\\

In conclusion,
brilliant multi-GeV CP (LP) $\gamma$-rays with polarization up to about 95\% are shown to be feasible in the nonlinear regime of Compton scattering with  ultrarelativistic  longitudinally (transversely) spin-polarized electrons. A photon number applicable for vacuum birefringence measurement in ultrastrong laser fields is achievable in a single-shot interaction. High degree of linear polarization is obtained due to the spin-dependent radiation reaction in a laser field with a small ellipticity, which induces separation of $\gamma$-photons with respect to the polarization.\\

 {\it Acknowledgement:} We thank J. Evers, T. Wistisen, Y.-S. Huang, R.-T. Guo and Y. Wang for helpful discussion. Y. Li, F. Wan and J. Li thank Prof. C. Keitel for hospitality.
 This work is supported by the National Natural Science Foundation of China (Grants Nos. 11874295, 11804269, 11905169), the National Key R\&D Program of China (Grant No. 2018YFA0404801) and the  Science Challenge Project of China (No. TZ2016099).

\bibliography{QEDspin}

\end{document}